\documentclass[10pt,letterpaper]{article}
\usepackage{opex3}

\begin{document}

\title{Dynamic saturation in semiconductor optical amplifiers: accurate model, role of carrier density, and slow light}

\author{Perrine Berger$^{1,2}$, Mehdi Alouini$^{1,3}$, J\'er\^ome Bourderionnet$^1$, \linebreak Fabien Bretenaker$^2$, and Daniel Dolfi$^1$}

\address{$^1$Thales Research \& Technology, 1 av. Augustin Fresnel, 91767 Palaiseau Cedex, France}
\address{$^2$Laboratoire Aim\'e Cotton, CNRS-Universit\'e Paris Sud 11, Campus d'Orsay, 91405 Orsay Cedex, France}
\address{$^3$Institut de Physique de Rennes, UMR CNRS 6251, Campus de Beaulieu, 35042 Rennes Cedex, France}

\email{Perrine.Berger@thalesgroup.com} 



\begin{abstract}  We developed an improved model in order to predict the RF behavior and the slow light properties of the SOA valid for any experimental conditions. It takes into account the dynamic saturation of the SOA, which can be fully characterized by a simple measurement, and only relies on material fitting parameters, independent of the optical intensity and the injected current. The present model is validated by showing a good agreement with experiments for small and large modulation indices.
\end{abstract}

\ocis{(250.5980)  Semiconductor optical amplifiers; (070.6020)  Continuous optical signal processing.} 


\section{Introduction}
\noindent The generation of continuously tunable optical delays is a
key element in microwave photonics. Among the targeted applications,
one can quote the filtering of microwave signals, the
synchronization of optoelectronics oscillators, and the control of
optically fed phased array antennas
\cite{Yao2009,Dolfi1996,Capmany2006}. With these applications in
view, large efforts are currently done in order to develop delay
lines based on slow and fast light effects
\cite{Chang-Hasnain2006,Su2006,Uskov2006,Pesala2008,Thevenaz2008}.
To date, one of the most mature approaches for integration in real
field systems is that based on Coherent Population Oscillations
(CPO) in semiconductor structures
\cite{Ku2004,Boula-Picard2005,Mork2005}. This approach offers
compactness, continuous tunability of the delay through injected
current control, and possible high-level parallelism
\cite{Maicas2007,Chen2009}. Obviously, the implementation of CPO
effects in microwave photonics delay lines relies on accurate
theoretical description of the underlying mechanisms in order to
develop reliable predictive models. Numerous theoretical models have
been developed in the past few years to describe CPO effects in
Semiconductor Optical Amplifiers (SOAs)
\cite{Agrawal1988,Zhou2007,Capua2008,Kim2008}. They are usually
based on a semi-classical description of the interaction between the
carriers and the input optical fields. These models offer a
comprehensive understanding of the gain saturation dynamics and
associated group index changes. However, on the one hand,
a complete model would require a detailed knowledge of the geometrical
and material parameters of the semiconductor structure
\cite{Chang2007,Connelly2001}. Unfortunately, most of them are
unknown especially when the SOA under consideration is a
commercially available device. On the other hand,
others, simpler, assumed that both the saturation power and
the carrier recombination lifetime are constant \cite{Su2006,Agrawal1988,Chen08,Mukai1990}. This assumption
applies when the SOA is operated at a fixed injection current
\cite{Haug1984,Rosencher2002}. However, the injection current and the input
optical power have to be tuned over a wide range in order to control
the speed of light into the SOA: consequently this assumption
restricts the predictive capability of a model describing
microwave-photonics delay lines using slow light in SOA.

In this paper we derive an improved model that enables to predict
the RF gain compression, the RF phase delay, and the
optical group delay and which is valid for all experimental
conditions for a given component. Furthermore, we show that the detailed knowledge of the inner geometrical and material
characteristics of the SOA is not required provided that some
preliminary and easy characterization measurements are conducted.
This model is then experimentally validated.

\section{Model} \label{part:theory}

We consider an optical carrier modulated by an RF signal and
injected in a traveling wave SOA. The total field is then composed
of the optical carrier of complex amplitude $E_0$ and two sidebands
of complex amplitudes $E_1$ and $E_2$. The total optical field E is normalized to include
the factor$\sqrt{\epsilon_0 n_0 c_0}$, i.e., the optical intensity is given
by \mbox{$I_{opt}(z,t)=\frac{1}{2}|E_{total}|^2 = U
+ M e^{-i \Omega t} + c.c.$}, under small RF signal
approximation. $U$ is the DC component of the
intensity, $M = \frac{1}{2} \left(E_0 E_2^* +E_1 E_0^*\right)$ is the beat-note term at the RF frequency $\Omega$.

The local equations for the propagation of the optical field
$E_{total}$ and the evolution of carrier density $N$ inside the SOA
are \cite{Agrawal1988}:

\begin{eqnarray}
\frac{dN(z,t)}{dt} &=&
\frac{I}{qV}-\frac{N(z,t)}{\tau_s}-\frac{g(z,t)|E_{total}(z,t)|^2}{\hbar
\omega}, \label{eq00} \\
\frac{d|E_{total}(z,t)|^2}{dz} &=& |E_{total}(z,t)|^2 \left[ -
\gamma + \Gamma g(z,t) \right], \label{eq0}
\end{eqnarray}
where $\gamma$ holds for the internal losses of the SOA,
$\Gamma g(z,t)$ is the material modal gain, $\tau_s$ is the carrier
lifetime, $I$ is the injected current, V is the volume of the active
region, and $\omega$ is the pulsation of the optical carrier $E_0$.
We introduce $N(z,t)=\bar{N}(z) + \Delta N(z) e^{-i \Omega t} +
c.c.$ and $g(z,t)=g(N(z,t))=\bar{g}(\bar{N}(z))+a(\bar{N}(z)) \Delta
N(z,t) e^{-i \Omega t} + c.c.$ where $a$ is the differential gain
$a(\bar{N})= \frac{\partial g}{\partial N} | _{\bar{N}}$. The
wavelength of the optical carrier is fixed. Consequently, the
equations Eq.~\ref{eq00} and Eq.~\ref{eq0} lead to:

\begin{eqnarray}
\frac{dU}{dz} &=& U \left[ - \gamma + \Gamma g(\bar{N}) \right], \label{eq1} \\
\frac{dM}{dz} &=& M \left\{ - \gamma + \Gamma g(\bar{N})\left(1-\frac{U/U_s(\bar{N})}{1 + U/U_s(\bar{N}) - i \Omega \tau_s(\bar{N})}\right) \right\}, \label{eq2}
\end{eqnarray}
where $U_s(\bar{N})$ is the saturation intensity defined as: $U_s(\bar{N})
= \frac{\hbar\omega}{a(\bar{N}) \tau_s(\bar{N})}$.

In most of the simple models, the common approach to
solve equations (\ref{eq1}) and (\ref{eq2}) is to consider $a$ and
$\tau_s$ constant with respect to the carrier density and thus over
the whole length of the device \cite{Su2006,Agrawal1988,Chen08,Mukai1990}. This approximation does however not
give account of strong saturation conditions, with high gain and
carrier density variations, which typically occur in quantum wells
structures with strong carrier confinement. In this paper, we
propose to consider the carrier density variation along the
propagation axis and its influence on $a$ and $\tau_s$. Our central
hypothesis is that $a$ and $\tau_s$ can be determined as functions
of the DC component of the optical intensity $U$ solely, allowing
these dependencies to be determined from gain measurements.

Let us first suppose that we fulfill the small signal condition. In
this case, the stimulated emission is negligible compared to the
spontaneous emission, leading to the unsaturated steady state
solution of the rate equation (Eq.~\ref{eq00}):
\begin{equation}\label{eq3}
    \frac{I}{q\ L\ S_{act}} = \frac{\bar{N}}{\tau_s},
\end{equation}
where $L$ is the length of the SOA, $S_{act}$ is the area
of the active section of the SOA. Moreover, we also
suppose in this case that the carrier density $\bar{N}$ is constant
along the SOA. These hypothesis are equivalent to consider
that the amplified spontaneous emission does not saturate the gain.
A verification of this assumption will be shown in section \ref{part:exp}. Under these
conditions, a measurement of the small signal modal gain $\Gamma
g_0$ versus $I$ will be equivalent, owing to Eq.~\ref{eq3}, to a
determination of the modal gain $\Gamma g$ versus $\bar{N}/\tau_s$.
Here, $\Gamma$ is the ratio $S_{act}/S_{guide}$ of the active to
modal gain areas in the SOA.

A last relationship between $\frac{\bar{N}}{\tau_s}$ and $U$ is then
required to determine the modal gain $\Gamma g$ as a function of
$U$. It is obtained by substituting $\Gamma
g(\frac{\bar{N}}{\tau_s})$ in the saturated steady state solution of
the carriers rate equation (Eq.~\ref{eq00}):
\begin{equation}
\frac{I}{q\ L\ S_{act}} - \frac{\bar{N}}{\tau_s} -
\frac{\Gamma g(\frac{\bar{N}}{\tau_s})}{\hbar
\omega} \frac{U}{\Gamma} = 0, \label{eq4}
\end{equation}
where the injected current $I$ is now fixed by the operating conditions.

Added to the previous relationship between $\Gamma g$ and $\frac{\bar{N}}{\tau_s}$, the Eq.~\ref{eq4} gives another expression of $\Gamma g$ as a function of $\frac{\bar{N}}{\tau_s}$, $\frac{U}{\Gamma}$ and $I$. Consequently, $\Gamma g$ and $\frac{\bar{N}}{\tau_s}$ can be known with respect to the local intensity $\frac{U(z)}{\Gamma}$ and the injected current $I$.

To solve Eq.~\ref{eq2}, we need to express $\bar{N}$ as a function of $\frac{U(z)}{\Gamma}$ and $I$. This is equivalent to express $\bar{N}$ with respect to $\frac{\bar{N}}{\tau_s}$ since $\frac{\bar{N}}{\tau_s}$ is known as a function of $\frac{U(z)}{\Gamma}$ and $I$.  Consequently, we model our SOA using the well-known equation \cite{Rosencher2002}:
\begin{equation}
\frac{\bar{N}}{\tau_s} = A \bar{N} + B \bar{N}^2 + C \bar{N}^3, \label{eq6}
\end{equation}
where $A$, $B$, and $C$, which are respectively the non-radiative, spontaneous and Auger recombination coefficients, are the only parameters that will have to be fitted from the experimental results.

Using Eq.\ \ref{eq6} and the fact that we have proved that $\bar{N}/\tau_s$ and $\Gamma g$ can be considered as function of $\frac{U(z)}{\Gamma}$ and $I$ only, we see that $\bar{N}$, $\Gamma a=\Gamma\frac{ \partial g}{\partial N}$, and $\frac{U_s}{\Gamma}= \frac{\hbar\omega}{\Gamma a \tau_s}$ can also be considered as functions of $\frac{U(z)}{\Gamma}$ and $I$. This permits to replace Eqs.~(\ref{eq1}) and (\ref{eq2}) by the following system:
\begin{eqnarray}
  \frac{d U}{dz} &=& U \left[ - \gamma + \Gamma g(\frac{U(z)}{\Gamma},I) \right], \label{eq7} \\
   \frac{dM}{dz} &=& M \left\{ - \gamma +  \Gamma g(\frac{U(z)}{\Gamma},I)\left[1-\frac{\Gamma U/U_s(\frac{U(z)}{\Gamma},I)}{1+\Gamma U/U_s(\frac{U(z)}{\Gamma},I)-i\Omega\tau_s(\frac{U(z)}{\Gamma},I)}\right] \right\}. \label{eq8}
\end{eqnarray}

Eqs. (\ref{eq7}) and (\ref{eq8}) are then numerically
solved: Eq.~\ref{eq7} gives $\frac{U(z)}{\Gamma}$, with the initial
condition $\frac{U(0)}{\Gamma}=\sqrt{\gamma_i}
\frac{P_{in}}{S_{act}}$, where $P_{in}$ is the optical input power. $\frac{U(z)}{\Gamma}$ can be then
introduced into Eq.~\ref{eq8}, and the microwave transfer function
of the SOA \mbox{$S_{21} = \gamma_i \frac{M(L)}{M(0)}$},
where $\gamma_i$ are the insertion losses, is then computed. If the
output power of the RF microwave signal is wanted, $P_{RF}=2 R
\eta_{ph}^2 \gamma_i |\frac{m P_{in} M(L)}{2}|^2$, with the initial condition $M(0) = 1$, where $m$ the input
modulation index, and $R$ and $\eta_{ph}$ are respectively the
photodiode resistive load and efficiency.

The microwave complex transfer function $S_{21}$ fully
characterizes the slow light properties of the SOA. Indeed the
optical group delay $\Delta \tau_g$ can be expressed as $\Delta
\tau_g(\Omega)=\frac{arg(S_{21}(\Omega))}{\Omega}$, and the group index $\Delta
n_g(\Omega)=\frac{c \ \arg(S_{21}(\Omega))}{L \ \ \Omega}$.

It is important to note that the recombination coefficients $A$, $B$
and $C$ are the only fitting parameters of our model. Once obtained
from experimental data, they are fixed for any other experimental
conditions. Moreover, the only geometrical required
parameters are the length $L$ of the SOA and the active area cross
section $S_{act}$. The derivation of a predictive model, independent
of the experimental conditions (current and input optical power) is
then possible, provided that the simple measurements of the total losses
and the small signal gain versus the current are conducted. The
above model lies in the fact that first, the spatial variations of
the saturation parameters are taken into account, and second, their
values with respect to the local optical power are deduced from a
simple measurement. These keys ideas lead to a very
convenient model of the microwave complex transfer function of the SOA,
and then of the slow light properties of the component. It can be
easily used to characterize commercial components whose design
details are usually unknown, as we will experimentally show in
the next section.

\section{Experiment} \label{part:exp}

\begin{figure}[t]
  \centering
  \includegraphics[width=12cm]{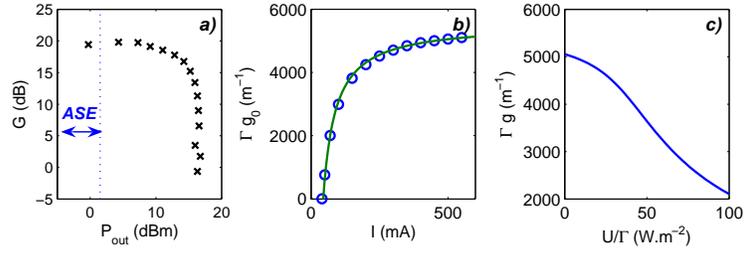}
  \caption{(a) Experimental fiber-to-fiber gain $G$ with respect to the output optical power $P_{out}$ at a strong
current (500 mA). The double arrow indicates the range of the measured ASE
output power. (b) Experimental small signal gain $\Gamma g_0$ as a function of the injected current $I$ at 1535 nm, and fitted by: \mbox{$\Gamma g_0 = C_1 - \frac{C_2}{I}$}, with \mbox{$C_1 = 5588.7 \mathrm{m^{-1}}$} and \mbox{$C_2 = 306.1 \mathrm{A^{-1}.m^{-1}}$}. (c) Deduced material modal gain $\Gamma g(U)$ as a function of the local intensity $U$ at 500 mA.}
  \label{fig:g0}
\end{figure}

In order to validate our model, we studied a commercially available
SOA (InP/InGaAsP Quantum Well Booster Amplifier from COVEGA). The length $L$ of this SOA is 1.50 mm and the active area
cross-section is set at $0.06 \ \mathrm{\mu m^2}$. We
proposed to compare the experimental and simulated complex transfer
function $S_{21}$ for a large set of operating conditions
($P_{in},I$). As explained in section \ref{part:theory}, the
study of the phase $arg(S_{21})$ is equivalent to the optical group
delay. We can then restrain our comparison to $S_{21}$. In order to fully characterize the response of the SOA
through $\Gamma g(U)$ as described in section \ref{part:theory}, the
preliminary step consists in measuring the total losses
and the unsaturated gain $\Gamma g_0(I)$ for different injected
currents.

\begin{figure}[t]
  \centering
  \includegraphics[width=7.8cm]{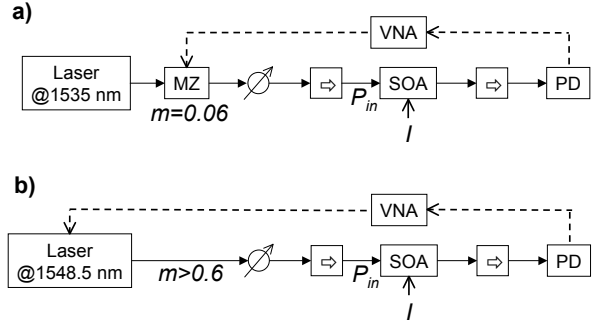}
  \caption{Experimental set-up. For small modulation index $m$, a laser is externally modulated by a Mach-Zehnder modulator (MZ) (a); for large modulation index, a directly modulated laser is used (b). In both cases, the input optical power $P_{in}$ is controlled through a variable optical attenuator; two optical isolators are used before and after the SOA. The photodetector (PD) restitute the RF signal. The Vector Network Analyser (VNA) is calibrated with the whole link without the SOA, in order to measure the RF transfer function of the SOA.}
  \label{fig:exp}
\end{figure}

\begin{figure}[h!]
  \centering
  \includegraphics[width=9cm]{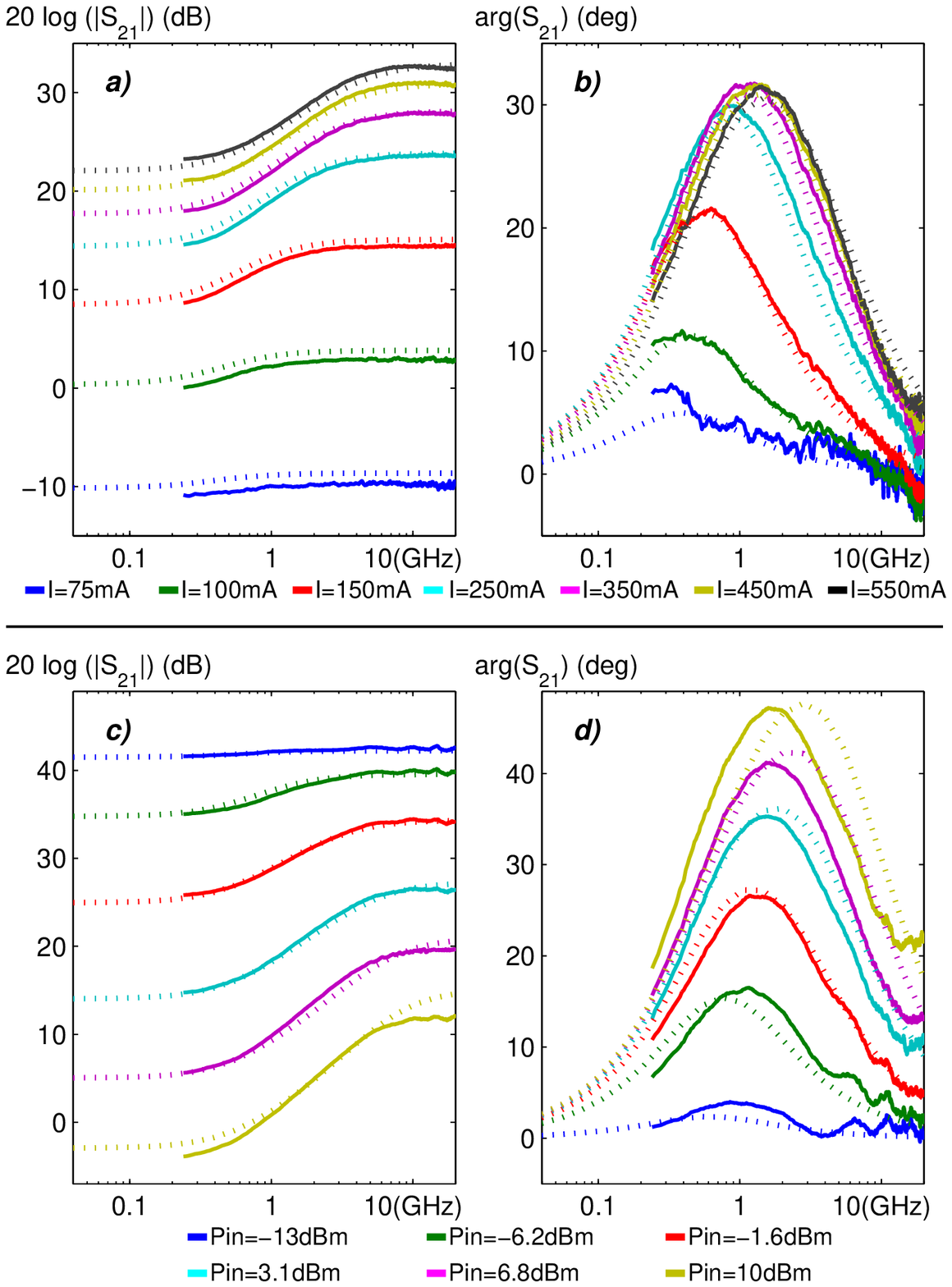}
  \caption{Low modulation index ($m=0.06$) : gain and phase shift simulations (dashed line) and experimental data (solid line) for (a) and (b): different injected currents at $P_{in}=0 \mathrm{dBm}$, and for (c) and (d): different optical input powers $P_{in}$ at $I=500 \mathrm{mA}$. The operating wavelength was $1535 \mathrm{nm}$.}
  \label{fig:faiblem_S21}
\end{figure}

The total losses are measured by the following experiment:
at low current, the output optical power is measured while a strong
input optical power is sent into the SOA. When the current is low
enough, the SOA is in the absorption regime: the resulting output power
is an increasing function of the input power (absorption saturation). When the current is above the transparency current, the resulting output power
becomes a decreasing function of the input power (gain saturation). Between
these two regimes, i.e. at transparency, the ratio between the output
power and the input power is exactly equal to the total losses. The total
losses of the SOA $\gamma_i \exp(- \gamma L)$ are measured to be equal to $-16.4 \mathrm{dB}$ in our case.

To measure only the unsaturated gain $\Gamma g_0$ despite
the amplified spontaneous emission, we measured the SOA unsaturated
RF gain $G_{RF}$ (=$|S_{21}|^2$) at a RF frequency $\Omega$ well
above $1/\tau_s$ (typically 20 GHz). The derivation of the modal gain $\Gamma g$ with respect to
$\frac{\bar{N}}{\tau_s}$ from the unsaturated gain $\Gamma g_0(I)$ is relying on the hypothesis that the
amplified stimulated emission (ASE) does not saturate the gain. In
Fig.~\ref{fig:g0}a, we represent the experimental fiber-to-fiber gain with
respect to the output optical power at a strong current (500 mA) and the
range of the experimental output power of the ASE. The maximum power of the ASE is equal to 1.54 dBm. Moreover, when the small signal measurement is performed, a maximum input optical power of $80 \ \mathrm{\mu W}$ was used, corresponding to an output optical power of 8.1 dBm for the maximum current. Consequently, both signal and ASE output power level are well below the output power required to saturate the gain (14.2 dBm for a 3dB gain reduction). Therefore, the experimental conditions match our preliminary assumptions. Under these conditions,
Eq.~\ref{eq7} can be simplified and integrated, leading to the
expression of the optical small signal gain $\Gamma g_0$:
$\exp(\Gamma g_0 L) = \frac{\sqrt{G_{RF}}}{\gamma_i \exp(- \gamma
L)}$. As shown in Fig.~\ref{fig:exp}a, we used a Vector Network Analyzer
(VNA) to measure the RF gain $G_{RF}$ for a small input power which
does not saturate the SOA (typically $10-80 \mathrm{\mu W}$). The
unsaturated gain of our SOA is displayed in Fig.~\ref{fig:g0}b. It
is empirically fitted by $\Gamma g_0 = C_1 - \frac{C_2}{I}$ with a
good agreement. From this simple measurement and using Eq.~\ref{eq4}, the material modal gain $\Gamma g$ is
then known as a function of the local intensity $U / \Gamma$ inside the
SOA (Fig.~\ref{fig:g0}c).

\begin{figure}[t]
  \centering
  \includegraphics[width=9cm]{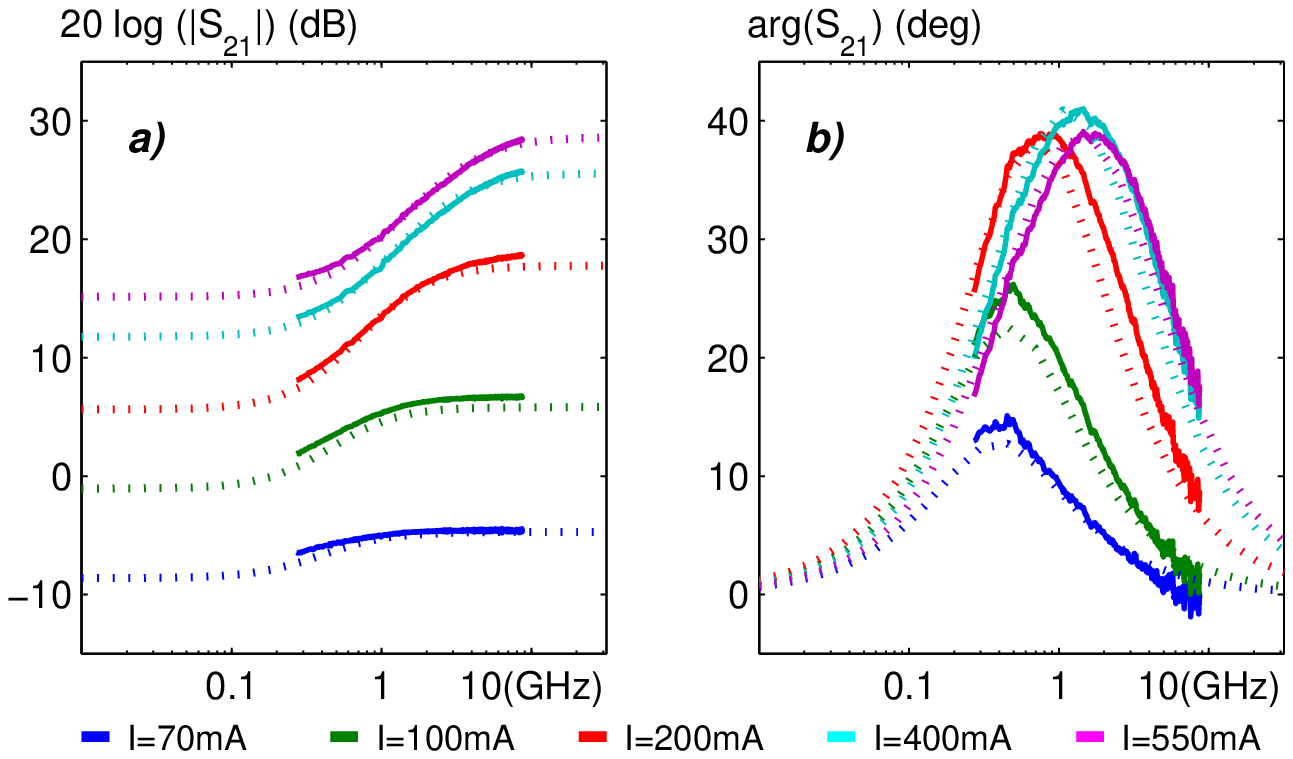}
  \caption{Large modulation index ($m>0.6$) : gain and phase shift simulations (dashed line) and experimental data (solid line) for (a) and (b): different injected currents at $P_{in}=0 \mathrm{dBm}$. The operating wavelength was $1548.5 \mathrm{nm}$.}
  \label{fig:fortm_S21}
\end{figure}
The complex RF transfer function of the SOA is measured thanks to a
VNA for small and large modulation indices (set-ups in
Fig.~\ref{fig:exp}). In Fig.~\ref{fig:faiblem_S21} and
Fig.~\ref{fig:fortm_S21}, we report the corresponding RF gains, $20
\log{|S_{21}|}$, and the measured evolution of the RF phase shift,
$\arg(S_{21})$, as a function of the modulation frequency $\Omega$.
In each of these figures, the plots labeled (a) and (b) correspond
to the evolutions of the RF gains and phase shifts versus RF
frequency, for different injected currents, while the plots labeled
(c) and (d) are obtained by managing the input optical power.

\bigskip

\section{Discussion} \label{part:dicus}
In Fig.~\ref{fig:faiblem_S21} and Fig.~\ref{fig:fortm_S21}, the
simulation results are reported in dashed line. The best fit values
for the recombination coefficients are: \mbox{$A= 2\times10^{9}
\mathrm{s^{-1}}$}, \mbox{$B=1.2\times10^{-10}  \mathrm{cm^3
s^{-1}}$}, \mbox{$C=1.8\times10^{-31} \mathrm{cm^6 s^{-1}}$}. These
values are in the range of what can be found in the literature for
semiconductor materials \cite{Ouacha93,Leu91,Bente08,Petrauskas92}.
The computed complex transfer function  shows a very good agreement
with the experimental data, both at small and large modulation
index, for any experimental conditions (injected current, input
optical power), and with a single set of the fitting parameters (A, B, C): our convenient model is predictive for any
experimental conditions.

\begin{figure}[t]
  \centering
  \includegraphics[width=11.5cm]{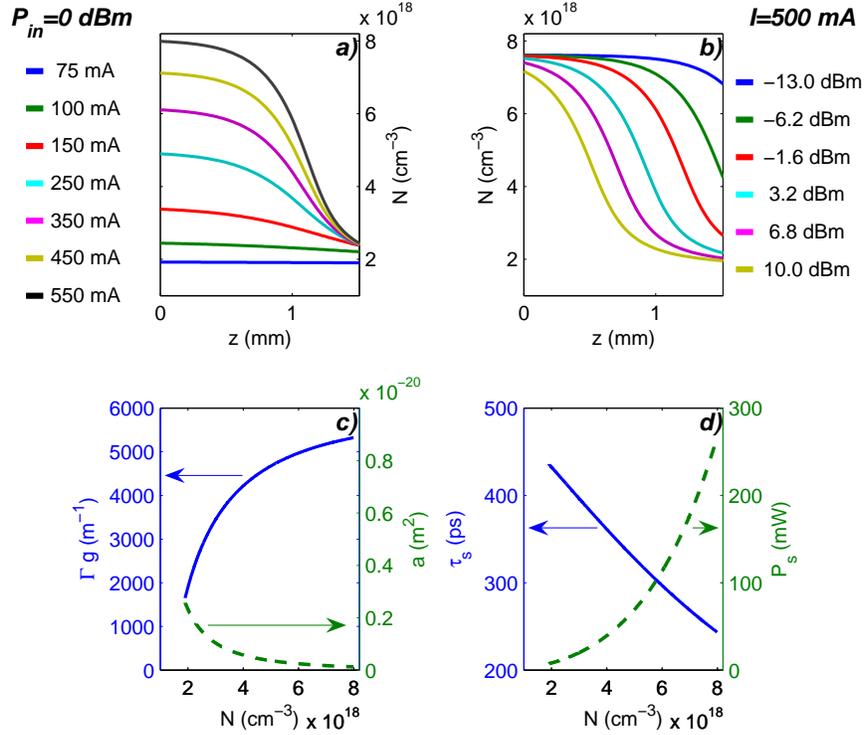}
  \caption{(a) and (b): Simulated carrier density $\bar{N}$ along the SOA: (a) at a fixed input optical power (0 dBm), for various currents; (b) at a fixed current (500 mA), for various input optical power. (c) and (d): Simulated variations with respect to the carrier density $\bar{N}$ of (c) the modal gain $\Gamma g$ (solid line), and the modal differential gain $a$ (dashed line); (d) the carrier lifetime $\tau_s$ (solid line), and the local saturation power $P_s$ (dashed line).}
  \label{fig:N_z}
\end{figure}

In order to highlight the weight of the spatial variations of the
carrier density and the saturation parameters, we plotted in Fig.~\ref{fig:N_z}a,\ref{fig:N_z}b the variations of the carrier density
$\bar{N}$ along the SOA for the different experimental situations of
Fig.~\ref{fig:faiblem_S21}. The subsequent variations of the modal
gain $\Gamma g$ and the saturation parameters $P_{s}$, $\tau_s$ and
$a$, with respect to $\bar{N}$, are displayed in
Fig.~\ref{fig:N_z}c,\ref{fig:N_z}d. We find at least one order of magnitude of
variation for almost all these parameters, which are nevertheless
often taken constant in literature for practical models \cite{Su2006,Agrawal1988,Chen08,Mukai1990}.  According
to Eq.~\ref{eq6}, this approximation can be justified when the variations
of $\bar{N}$ along the SOA are relatively not too strong, that is for moderate bias current ($<150 \ \ \mathrm{mA}$ in our case) or a high bias current, but low optical power. However, for any other condition, and especially in the case of quantum well
or quantum dots structures, it is necessary to take into account the
saturation dynamics along the propagation to ensure good
performances of the model and robustness versus changes in
experimental conditions. Indeed, Fig.~\ref{fig:N_z} shows that considering $P_{s}$, $\tau_s$ and $a$
constant, and then $\Gamma g$ linear with $\bar{N}$, drastically
limits the range of experimental conditions $(P_{in}, I)$ where such
models are valid, which forces the saturation parameters to be
adjusted with the current and/or the optical input power.

Our improved model is still easy to use, even for
commercial components, but despite the hypothesis we were compelled
to make, it remains valid for a large range of experimental
conditions, with a reduced set of unknown - and thus fitted-
parameters. These advantages have been achieved by taking into
account the spatial variation of the saturation parameters and by
showing that their values as a function of the local optical power
can be retrieved from a simple measurement. It ensures that the
model only relies on material fitting parameters, independent of the
optical intensity and injected current.

The slow light properties are then also modeled for a large
range of the input optical powers $P_{in}$ and injected currents
$I$, which is essential from the operational point of view, since the speed of light in SOA is controlled by
these two key parameters. While the applications of slow light in
SOA are taking shape, a convenient and accurate model with the
parameters tuning the delays is a necessary tool to fully
characterize the effect of slow light in SOA on a microwave link, or
to develop new architectures improving the slow light properties. This
model could be easily used when an optical filtering is performed
after the SOA to enhance the slow light effect, as described in
\cite{Chen08,Shumakher09}. In this case, Eq.~\ref{eq8} just has to
be replaced by the corresponding coupled equations in $E_1$ and
$E_2$. Moreover, to take into account higher order coherent
population oscillations \cite{Duill09}, the present model can be
generalized using equations similar to Eq.~\ref{eq8} for each
harmonic of the optical intensity. The determination of $\Gamma g$
as a function of $U$ is slightly more subtle in this case: it is
presented in another paper, in order to study the harmonic
generation and the intermodulation products \cite{OE_arxiv}.

\section{Conclusion}
We developed an improved but still convenient model in
order to predict the RF behavior and slow light properties of the
SOA, valid for any experimental conditions (input optical power,
injected current). It takes into account the spatial variations of
the saturation parameters along the SOA, which are fully
characterized by the simple measurement of the small signal gain.
The resulting model only relies on material fitting parameters,
independent of the optical intensity and injected current. We showed
a remarkably good agreement between the model and the experimental
data, at small and large modulation indices. The ease of use and the
accurate prediction obtained for any experimental conditions will be
useful to characterize the effect of slow light in SOA on a
microwave link, and to develop new designs improving the slow light
properties. The key ideas of this improved model can easily be used when optical
filtering is performed after the SOA. A generalization of our
approach will be carried out in a next step, in order to determine
the harmonic generation, intermodulation products and spurious free
dynamic range, for a full characterization of a SOA based
opto-electronic link.

\section*{Acknowledgments}
The authors acknowledge the partial support from the "D\'{e}l\'{e}gation G\'{e}n\'{e}rale pour l'Armement" DGA/MRIS and from the GOSPEL EC/FET
project.
\end{document}